%% file: main.tex
\documentclass[sigconf]{acmart}

\acmConference{Technical Report}{August, 2024}{Porto, Portugal}

\usepackage{hyperref}
\usepackage{subcaption}

\usepackage[normalem]{ulem}

\newcommand{\ignore}[1]{}

\begin{document}

\title[Enhancing Adaptive Mixed Criticality with Deep Reinforcement Learning]{Enhancing Adaptive Mixed-Criticality Scheduling\\ with Deep Reinforcement Learning\\}

\author{Bruno Mendes}
\affiliation{%
  \institution{Department of Informatics Engineering (DEI), Faculty of Engineering of the University of Porto (FEUP)}
  \city{Porto}
  \country{Portugal}
}
\email{up201906166@fe.up.pt}

\author{Pedro F. Souto}
\affiliation{%
  \institution{Department of Informatics Engineering (DEI) and CISTER Research Centre, Faculty of Engineering of the University of Porto}
  \city{Porto}
  \country{Portugal}
}
\email{pfs@fe.up.pt}

\author{Pedro C. Diniz}
\affiliation{%
  \institution{Department of Informatics Engineering (DEI), Faculty of Engineering of the University of Porto (FEUP)}
  \city{Porto}
  \country{Portugal}
}
\email{pedrodiniz@fe.up.pt}

\renewcommand{\shortauthors}{Mendes et al.}

\begin{abstract}
Adaptive Mixed-Criticality (AMC) is a fixed-priority preemptive scheduling algorithm for
mixed-criticality hard real-time systems. It dominates many other 
scheduling algorithms for mixed-criticality systems, but does so at the
cost of occasionally dropping jobs of less important/critical tasks, when low-priority jobs overrun their time budgets. 
In this paper we enhance AMC with a deep reinforcement learning (DRL) approach based on a Deep-Q Network. 
The DRL agent is trained off-line, and at run-time adjusts the
low-criticality budgets of tasks to avoid budget overruns, while ensuring that no job misses its deadline if it does not overrun its budget.
We have implemented and evaluated this approach by simulating
realistic workloads from the automotive domain. The results 
show that the agent is able to reduce budget overruns by at least up to $50$\%, even when the budget of each task
is chosen based on sampling the distribution of its execution
time. To the best of our knowledge, this is the first use of DRL
in AMC reported in the literature.
\end{abstract}


\maketitle

\input{introduction}
\input{related}
\input{background}
\input{approach}
\input{agent_pedro}
\input{environment}
\input{evaluation}
\input{conclusions}


\bibliographystyle{ACM-Reference-Format}
\bibliography{references}

\end{document}

%% file: introduction.tex
\section{Introduction}

Many systems in the automotive and avionics industries are 
designed as mixed-criticality systems (MCS), i.e. as
systems with tasks with more than one level of criticality that execute on the same hardware platform. 
From the point of view of scheduling, the overriding goal is to ensure a high level of assurance regarding timing of critical tasks, while leveraging the increasingly 
available computing power for lower criticality tasks. Sharing resources is also a goal to reduce system complexity, energy/power and weight, in particular in airborne systems where it
is at a premium.

Adaptive Mixed Criticality (AMC) \cite{baruah_response-time_2011} is a fixed-priority preemptive scheduling algorithm, that descends from Vestal's seminal scheduling algorithm for MCS~\cite{vestal_preemptive_2007}. 
In this approach, the tasks are assigned different criticality levels: HI and LO, in its simplest form. In this case, the system may operate in two modes, HI and LO. The system starts in LO-mode. In LO-mode, every task may execute for its (time) budget, which is fixed and assigned off-line. Whenever a job exceeds its budget, the system changes to HI-mode and no new LO-tasks are released in that mode.

While effective, this approach has a major drawback. 
Just because a task has a low criticality, it does
not mean that it is less important and that it can be abandoned \cite{esper18_academic_understanding_MCS, ernst16_mcs_misconceptions_I3E_DT}.

In this paper, we propose a novel approach based on deep reinforcement learning (DRL) to adjust the budgets of both LO- and HI-criticality tasks at run time, so as to reduce the number of budget overruns, and therefore the number of LO-criticality job cancellations.
The DRL agent is trained off-line, so as to reduce the load on the system at run-time. At run-time, it takes as input the current budgets and execution times of the different jobs, and outputs the adjustments to apply to the budgets of different tasks. We ensure that in spite of these changes the system is still schedulable according to AMC-rtb~\cite{baruah_response-time_2011}, a sufficient schedulability test for AMC that requires the solution of recurrences, without running AMC-rtb at run-time.
To avoid interference with application tasks, the agent runs with the lowest priority as a high-criticality task, thus ensuring that it can adjust the budgets after a switch from LO- to HI-mode.

The main contributions of this paper are as follows:
\begin{itemize}
\item Describes a DRL agent using a Deep-Q Network that adjusts at run-time the budgets of tasks in LO-mode so as to reduce time budget overruns, while providing the same guarantees as AMC: i) all HI jobs meet their deadlines; ii) all LO tasks that do not overrun their budgets meet their deadlines;
\item Describes an event-driven simulator that is able to simulate the timing behavior of realistic automotive applications~\cite{kramer_real_automotive} on a uniprocessor platform;
\item Presents an exhaustive experimental evaluation of the DRL agent for a realistic tasks workload drawn from an automotive application~\cite{kramer_real_automotive}.
\end{itemize}

The results we obtained show that the DQN-agent can reduce the number of budget overruns of both LO-criticality and HI-criticality jobs by at least 50\%.

The remainder of this paper is organized as follows. In the next section, we discuss the related work. In Section~\ref{sec:background} we provide some background on AMC-rtb, and on deep reinforcement learning.  Section~\ref{sec:approach} describes the proposed approach. The design of the DRL agent is presented in Section~\ref{sec:agent}. Section~\ref{sec:environment} describes the simulator used for training and evaluating the agent. In Section~\ref{sec:evaluation} we evaluate the proposed approach using a scenario-based simulation. Section~\ref{sec:conclusion} concludes with a summary.

%% file: related.tex
\section{Related work}

AMC~\cite{baruah_response-time_2011} not only dominates many other fixed-priority preemptive scheduling (FPPS) for mixed-criticality systems, but it also has best performance for fixed priority sporadic schemes on uniprocessor systems~\cite{huang2014_impl_eval_mc_schemes_tecs}.
A key feature of AMC is that it may drop LO-criticality tasks when the HI-criticality
tasks experience overload. This can be seen as a shortcoming as just because a 
task has a low criticality, it does not mean that it is less important and that it can be 
dropped. To mitigate this issue, researchers have developed several approaches 
which can be  broadly classified in three classes: modifications to the model, 
design-time solutions and run-time solutions.

\paragraph{Model changes} These solutions avoid the issue, by allowing lower 
criticality tasks to run even upon a mode change to a higher priority level, but 
reduce their load to compensate for the excess demand by higher-criticality tasks. 
As pointed out in~\cite{burns13_more_pratical_model_for_MCS_WMC}, in the context
of FPPS, this can be achieved by one of the 2 following approaches: 1) reduce 
the execution time of lower-criticality tasks; and 2) increase the periods of 
lower-criticality tasks. A third approach based on changing priorities does no
really reduce the demand, but because lower criticality tasks will execute with 
lower priority they will not interfere with higher priority tasks.

Most research following this approach has focused on reducing the execution time 
of lower-criticality tasks. The work in \cite{gettings15_mcs_weakly-hard_rtns} combines 
AMC with weakly-hard constraints, more specifically it allows LO-tasks to skip up to $s$
deadlines out of $m$ consecutive deadlines. Other authors ({\em e.g.} \cite{davis22_compensating_AMC_RTNS})
proposes the {\em compensating AM} model, which assumes that low-criticality tasks have
imprecise versions, which require less execution time than their more complete versions,
and that are executed when the system is in high-criticality mode, to ensure all jobs meet
their deadlines regardless of their criticality and of the mode of operation, as long as they
do not exceed their budgets.

An alternative way to reduce the demand by LO-tasks, is to drop the jobs of a subset of 
the LO-tasks~\cite{huang13_interference_constraint_graph_etfa, flemming14_notion_of_importance_in_MCS_WMC}.
Santinelli and Guo~\cite{santinelli18_sensitivity_analysis_for_MC_etfa} extend the usual mixed-criticality
model by defining the system criticality mode as a combination of the task criticality modes, and provide
a schedulability analysis based on bound functions.
Sundar and Easwaran \cite{sundar19_practical_degradation_model_for_MCS_isorc} propose a context aware approach to choosing
which tasks to degrade rather than their criticality level, {\em i.e.,} they allow for reducing the budgets of high criticality jobs to avoid dropping lower criticality jobs. As a consequence, modes of operation are not directly related to criticality levels.
Burns and Davis \cite{burns20_AMC_semi-clarivoyance_rtss} extend AMC to the semi-clairvoyance model~\cite{agrawal19_semi-clairvoyance_MCS_rtss}, which assumes that upon arrival a job can indicate whether it will overrun its budget. Zhang et al.~\cite{zhang24_AMC_with_semi-clairvoyance_graceful_degradation_tecs} extend the Compensating-AMC model with semi-clairvoyance.
In yet another approach, Baruah and Ekberg~\cite{baruah21_graceful_degradation_semi-clairvoyant_ecrts} 
study the graceful degradation of semi-clairvoyant scheduling.

The approach proposed in this paper is orthogonal to virtually all these model changes, and therefore can be combined with them to improve their results.

\paragraph{Design-Time changes} These changes assume the AMC model and do not imply changes to its scheduling algorithm at run-time. Instead, they propose different parameter settings from those proposed in~\cite{baruah_response-time_2011}.

Santy {\em et al.}~\cite{santy12_relaxing_MCS_strictness_sensibility_analysis_ecrts} and later Burns and Baruah~\cite{burns13_more_pratical_model_for_MCS_WMC} (for
the Compensating AMC model)
use sensitivity analysis to increase the budgets of 
runnable tasks in each criticality level. 
This way, the likelihood that a budget overrun will
occur will be lower, consequently reducing the 
likelihood of the system changing to a higher criticality mode.  
Similarly, Santinelli and Guo~\cite{santinelli18_sensitivity_analysis_for_MC_etfa} apply sensitivity 
analysis to explore trade-offs between schedulability and criticality levels, in their proposed model.

\paragraph{Run-time changes} These approaches assume
one of the models described above and require changes
to the scheduling algorithm.
These approaches are the closest to our work, which also
requires minor changes to the implementation of AMC scheduling; namely the possibility of adjusting task
budgets at run-time, and also the addition of a component, the DRL agent, that executes at run-time and computes
the budget adjustments.

Bate {\em et al.}~\cite{bate15_bailout_protocol_ecrts} use accounting of budget under and overruns in HI-criticality mode to reduce the time in HI-mode, during which no newly released LO-criticality jobs are executed.
Furthermore, they use sensitivity analysis to compute
the budgets in LO-mode, so as to make changes to HI-mode
less likely. 
Furthermore, in~\cite{bate17_enhanced_bailout_protocol_I3E_TSE} the authors enhance the bailout protocol by reclaiming unused time when a task executes for less than its worst-case execution time budget.
Iacovelli and Kirner~\cite{iacovelli19_lazy_bailout_approach_designs} extend the bailout protocol with a separate priority queue for LO-priority jobs that would be dropped in degraded mode, a sub-mode of HI-criticality mode, for execution when the system is idle.
Papadopoulos {\em et al.}.~\cite{papadopoulos18_control_theoretic_approach} propose adjusting budgets at run-time as we do. However, their solution differs from ours on two accounts. First, they use a control-theoretic approach, second they assume a server-based approach.
Massaro {\em et al.}~\cite{massaro2018_integrating_proactive_mode_changes_arXiv} use a Single Layer Feedforward Network (SLFN) with a Kalman filter for predicting a budget overrun and proactively change the operating mode.
Lastly,  Hu {\em et al.}~\cite{hu19_online_mode-switch_procrastination_in_MCS_rts} use monitoring of task execution times and update a common overrun budget shared among all tasks. This allows to postpone mode-switch as long as there is enough unused capacity to compensate for budget overruns.

In this review of related work, we focused mainly on work to reduce service degradation of LO-criticality tasks in the Adaptive Mixed Criticality, {\em i.e.} fixed-priority preemptive scheduling, for uniprocessor system. For a more detailed review of the research on mixed criticality systems in the wake of Vestal's seminal work~\cite{vestal_preemptive_2007}, please read Burns and Davis thorough survey~\cite{burns17_survey_research_mcs_ACM-CS}. A more recent version of this survey can be found in \cite{,burns22_MCS_review_TR}.

%% file: background.tex
\section{Background}\label{sec:background}

\subsection{AMC Scheduling}\label{sec:AMC}

\input{AMC.tex}

\subsection{Reinforcement Learning}

Reinforcement learning (RL) aims to determine the optimal behaviour for an agent interacting with an environment. RL problems typically fit the definition of a Markov Decision Process (MDP) in that they are modelled by a set of states \(S\), a set of actions \(A\), a reward function \(R: S \times A \rightarrow \mathbb{R}\) and a state transition function \(T: S \times A \times S \rightarrow [0,1]\).

In an RL framework, the goal of an agent is to find a policy \(\pi\), a mapping of states to actions (\ref{eq:rl_policy}), that maximizes the sum of obtained rewards in the long term.
\begin{equation} \label{eq:rl_policy}
    a_t \sim \pi_{\theta}(s_t)
\end{equation}

\subsubsection{Traditional learning methods}

Training an agent in an RL problem is adjusting the policy throughout interactions with the environment, usually called \textit{episodes}.

Classical, value-based RL algorithms are \textit{temporal-difference methods} since updates to the policy are performed according to the difference in the measured future value and the estimated future value \cite{modern_valuebased_rl} of states. \textit{Q-learning}, an example of value-based RL, works by building a value function \(Q(s,a)\) that is stored in a table of finite size \(S \times A\) and updated iteratively with a learning rate \(\alpha\) and new values calculated using the Bellman equation (\ref{eq:qlearning_update}).
\begin{equation} \label{eq:qlearning_update}
    Q_{new}(s, a) \leftarrow (1 - \alpha) \cdot Q(s, a) + \alpha \cdot \left( r + \gamma \cdot \max_{a'} Q(s', a') \right)
\end{equation}

where \(s'\) is the state if one takes action \(a\) in state \(s\) and \(r\) is the reward by taking action \(a\) in state \(s\).

The main issue with Q-learning and other traditional tabular-based RL algorithms is that they can't work with large state representations due to memory limitations. In the real-time field, that issue is even more critical.

\subsubsection{Deep Q-Network}

The Deep Q-Network (DQN) algorithm replaces the Q-value function of Q-learning with a deep neural network, the \textit{policy network}, that outputs the Q-values of available actions in a state given as input.

Training a DQN-based agent requires using an auxiliary network, the \textit{target network}, a snapshot of an older version of the policy network. While performed actions are always provided by the output of the policy network, losses are computed against the output of the target network to stabilize training. Similarly, the target network is fed with a random sample of prior actions, saved in \textit{replay memory}, instead of the most recent action to break the correlation between states.

%% file: AMC.tex
Adaptive mixed criticality (AMC) was proposed by Baruah et al. in \cite{baruah_response-time_2011} as a fixed-priority scheduling scheme for mixed criticality systems that takes advantage of the ability of a platform to monitor at run-time the execution time of all jobs to improve system schedulability.

\subsubsection{System model}
The system is composed of a set of sporadic tasks. Each task, $\tau_i$ is characterized by a tuple $(T_i, D_i, L_i, \overset{\to}{C_i})$, where $T_i$ is the period, i.e. the minimum time between the arrival of consecutive jobs of the task, $D_i$ is the deadline (relative to the arrival time), $L_i$ is its criticality level, either HI or LO (HI $>$ LO), and $\overset{\to}{C_i}$ is the worst-case execution time (WCET) vector, with one WCET estimate per criticality level equal or lower than $L_i$. If $L_i$ is HI, then $C_i(HI)\ge C_i(LO)$ (otherwise, $C_i(HI)$ is not defined).

We assume that, regardless of the task level, the WCET estimate for level LO may not be safe, i.e. may be smaller than the actual WCET of the task. On the other hand, we assume that the WCET estimate for level HI is safe.

With respect to the platform, we assume that it has a single core and that it provides support for monitoring at run-time the execution time of every job of all tasks. This functionality is commonly available on many safety-critical real-time platforms \cite{baruah_response-time_2011}.

AMC can be extended to systems with a larger number of criticality levels. For ease of presentation, we consider only two criticality levels. In addition, we refer to LO (HI) criticality level tasks as LO-tasks (HI-tasks), and similarly to their jobs, i.e  LO-jobs (HI-jobs). Likewise we refer to the WCET estimate for LO (HI) criticality level as LO-WCET (HI-WCET).

\subsubsection{Scheduling algorithm}

AMC uses a fixed-priority algorithm, i.e. it assumes that each task has a unique priority and that these priorities are totally ordered. However, this algorithm operates in two modes, LO and HI. For conciseness we use LO-mode (HI-mode) to refer to mode LO (HI). 

In both modes, tasks are scheduled according to their priorities. 

In LO-mode, every task can run and is assigned a budget equal to its LO-WCET.  However, in HI-mode, only HI-tasks can run. Given that we assume that a HI-task never exceeds its HI-WCET, it is irrelevant to assign a budget to a HI-task in HI-mode (and also to a LO-task, which never runs in HI-mode). Thus, budgets are used only in LO-mode, and we refer to them using the word budget without a mode qualifier.

If a job exceeds its budget, all LO-jobs are cancelled and the scheduler switches to HI-mode.

The scheduler reverts to LO-mode when there are no jobs ready to run, i.e. when the processor is idle.

\subsubsection{Schedulability Analysis}

AMC-rtb is a sufficient response-time analysis for AMC introduced in \cite{baruah_response-time_2011}. It is an extension of fixed-priority response-time analysis \cite{joseph_finding_rts}, which computes an upper-bound on the response-time, $R_i$, of each task $\tau_i$. If for every task $\tau_i$, $R_i \le D_i$, then the task set is deemed schedulable. AMC-rtb comprises 3 analyses, one for LO-mode, another for stable HI-mode and another one for mode switch. 

LO-mode analysis is just an application of standard fixed-priority response-time for a task set composed of all the tasks in the system, and assuming that the WCET of each task is $C_i(LO)$, i.e. its budget. More specifically, the response time of task $\tau_i$ in LO-mode, $R_i^{LO}$ is given by the solution to the following recurrence:
\begin{equation}
    R_i^{LO} = C_i(LO) + \sum_{j \in \mathbf{hp(i)}} \left\lceil \frac{R_i^{LO}}{T_j} \right\rceil C_j(LO) \label{eq:amc-rtb-lo}
\end{equation}
where $\mathbf{hp(i)}$ is the set of all tasks with priority higher than that of task $\tau_i$. Response time analysis for stable HI-mode is similar, but applies only to HI-tasks.

The recurrence for determining the response time of a HI-job that is caught by mode switch, i.e. that arrived before the mode switch and completes only after mode switch, $R_i^*$, is:
\begin{align}
      R_i^*  = C_i(HI) + \sum_{j \in \mathbf{hpL(i)}} \left\lceil \frac{R_i^{LO}}{T_j} \right\rceil C_j(LO)  + \sum_{j \in \mathbf{hpH(i)}} \left\lceil \frac{R_i^*}{T_j} \right\rceil C_j(HI) \label{eq:amc-rtb-*} 
\end{align}
where $\mathbf{hpL(i)}$($\mathbf{hpH(i)}$) are the set of LO-tasks(HI-tasks) with priority higher than that of task $\tau_i$. Note that the second summation upper-bounds the interference by higher-priority HI-tasks, whereas the first summation upper-bounds the interference by higher priority LO-tasks. In the latter, we use $R_i^{LO}$ rather than $R_i^*$ as the numerator, because a mode change can occur at the latest $R_i^{LO}$ after the releaase of the job: if it does not, the job must complete before $R_i^{LO}$ and therefore will not be caught by the mode change. Note that, in AMC-rtb, the response time analysis for mode switch subsumes the response time analysis for stable HI-mode, thus there is no need for the latter.

A tighter response time analysis, AMC-max, that takes into account the mode switch instant is also provided in \cite{baruah_response-time_2011}. Because, this analysis has higher computational complexity, which is relevant as we shall discuss below, and several works, e.g. \cite{fleming13_extending_wmc}, report that the improvement achieved by AMC-max is relatively small, we will not present AMC-max. 

\paragraph{Priority assignment} AMC can use any priority assignment policy. It has been shown \cite{vestal_preemptive_2007} that deadline monotonic is not optimal, but, as argued in \cite{baruah_response-time_2011}, AMC-rtb satisfies the necessary conditions \cite{davis11_improved_priority_assignment_rts} for the optimality of Audsley algorithm \cite{audsley01_optimal_fixed_priority_assignment}. Furthermore, the number of tests to determine the optimal priority assignment is upper-bounded by $(2n - 1)$ \cite{baruah_response-time_2011}, rather than $n(n+1)/2$, the upper-bound in the general case, where $n$ is the number of tasks.

%% file: approach.tex
\section{Approach}\label{sec:approach}

AMC with AMC-rtb dominates many other fixed-priority mixed criticality algorithms and respective schedulability tests \cite{baruah_response-time_2011}. However, it does so at the cost of not running LO-jobs until the system becomes idle, when some job overruns its budget. Even though this behavior should not put the system at risk (otherwise, the respective task should not be classified as a LO-task), it may certainly affect the quality of service of the system.

To reduce budget overruns by jobs in the system, we propose to use a deep reinforcement learning agent, which will adjust the budgets of all the jobs at run-time. The agent is trained off-line, i.e. prior to deployment.

To minimize the interference on application tasks, the agent runs as a task with a priority lower than that of any application task. This also ensures that when the agent terminates, no other task is ready to execute, i.e. the system is idle, and therefore we can apply its output, i.e. the proposed adjustments to the budget. 

There is however an issue. In AMC, a task budget is its LO-WCET, which is used both in \eqref{eq:amc-rtb-lo} and \eqref{eq:amc-rtb-*}, AMC-rtb response-time recurrences. Thus, when the task budgets change, the worst case response time of some tasks may exceed the value computed by AMC-rtb. In particular, the response time of HI-jobs caught by a mode change may exceed their deadlines. To prevent this from happenning the output of the agent is validated before applying it to the system. The budget changes are applied only if it is guaranteed by AMC-rtb that no HI-job will ever exceed its deadline when running AMC with the new \emph{(budget) configuration}, i.e. the set of task-budget pairs.

Running AMC-rtb at run-time can be time-consuming given the need to solve two recurrences per HI-task: one to compute $R_i^{LO}$ and the other to compute $R_i^{*}$. Therefore, the agent only ensures that the budget changes do not invalidate the analysis done at design-time. More specifically, it checks that for every HI-task $\tau_i$,
\begin{align}
B_i + \sum_{j \in \mathbf{hp(i)}} \left\lceil \frac{R_i^{LO}}{T_j} \right\rceil B_j & \le R_i^{LO} \label{eq:amc-val-hi-lo} \\
C_i(HI) + \sum_{j \in \mathbf{hpL(i)}} \left\lceil \frac{R_i^{LO}}{T_j} \right\rceil B_j 
                + \sum_{j \in \mathbf{hpH(i)}} \left\lceil \frac{D_i}{T_j} \right\rceil C_j(HI) & \le D_i \label{eq:amc-val-*} 
\end{align}
where $R_i(LO)$ is the response time of $\tau_i$ computed at design time using AMC-rtb, $B_k$ is the budget of task $\tau_k$. These tests are safe in the sense that if the set of budgets $B_k$ satisfies these inequalities, then for each HI-task the worst case response time computed by AMC-rtb is still an upper-bound of the response time when AMC uses that configuration. There are however configurations that fail theses tests, but that would nevertheless satisfy $R_i^* \le D_i$ using AMC-rtb. Thus, the cost of not running AMC-rtb at run-time is that some valid configurations will not be applied because they are deemed invalid.

For some configurations validated by these tests, condition \eqref{eq:amc-val-hi-lo} may not hold for some LO-tasks. Thus, for LO-tasks we check that:
\begin{equation}
B_i + \sum_{j \in \mathbf{hp(i)}} \left\lceil \frac{D_i}{T_j} \right\rceil B_j  \le D_i \label{eq:amc-val-lo} 
\end{equation}
If these additional checks represent a significant load for the agent, the agent may skip them. In this case, we rely on the agent ability to select the right configuration that prevents jobs from exceeding their budgets and therefore their deadlines. However, there is a possibility that increasing some budgets will reduce the number of budget overruns, but increase the number of deadline misses. Adding condition \eqref{eq:amc-val-lo} for each LO-task, would exclude configurations that may lead to deadline misses by LO-tasks.

It is possible to reduce the average run-time overhead of these checks, by limiting the number of tasks that are checked. As explained in the next section, every time the agent runs it modifies the budgets of only a subset of the tasks. Thus rather than checking the validity for all tasks, the agent can check the validity only for those tasks that have higher or equal priority to those with modified budgets.

Finally, we observe that all the ceilings in conditions \eqref{eq:amc-val-hi-lo}, \eqref{eq:amc-val-*} and \eqref{eq:amc-val-lo} can be precomputed at design time. Therefore, executing these checks requires only sums and sums of products, which are common operations in deep learning algorithms and, therefore, have highly optimized implementations.

%% file: agent_pedro.tex
\section{Agent design}\label{sec:agent}

To design the RL agent, we formulate the problem as an MDP. In particular, we need to identify the environment, to specify its state representation, to identify the set of possible actions and define the rewards.

Given our problem, the environment is composed of the scheduler, the task set and the platform. 

\subsection{State}

The state is one of the two inputs of the agent during training and the sole input at run-time. This means that at run-time the agent must be able to take appropriate actions, i.e. adjust the task budgets so as to reduce budget overruns, based only on the state.

Thus we represent the state of an environment with a task set $\mathcal{T} = \{\tau_1, \tau_2, \ldots, \tau_n\}$ as a sequence of tuples $s = \{(B_1, c_1), \ldots, (B_n, c_n)\}$, where $B_i$ is the budget of task $\tau_i$ and $c_i$ is the execution time of its most recent job. The values are normalized using a min-max scale, where the minimum and maximum are the best-case execution time (BCET) and the worst-case execution time (WCET) of the corresponding task, both of which are computed as described in Section~\ref{sec:evaluation}.

\subsection{Action set}

As described above, the set of actions are changes to the budget of the different tasks. 

Due to the nature of DQN, the action set needs to be finite: the output layer of the DQN has one node per action. In order to limit the size of the DQN, we decided that each action should increase the budget of one task and decrease the budget of two other tasks. Thus, the number of outputs of the DQN is given by: $n(n-1)(n-2)/2$, where $n$ is the number of tasks in the task set. In order to avoid sudden changes to budgets and also to keep the processor utilization in LO-mode roughly constant, budget increases are of 10\% whereas budget reductions are of 5\%. 

Note that the DQN is oblivious to the semantics of the actions it outputs. It learns by consistently applying the action associated with the output with a maximum value, and receiving the corresponding rewards.

\subsection{Rewards assignment}

The agent needs to receive feedback for its actions during the DQN training phase. These rewards are important for the DQN to adjust the weights of the edges between its nodes, improving the quality of its decisions. That is done by computing the losses to the target network output using the mean-squared error and propagating it backwards in the policy network. Finally, they are optimized using an Adam optimizer.

We assign rewards to scheduling events depending on their desirability for the quality of the scheduling of the system:
\[
\text{reward(event)} = 
\begin{cases}
    0.1 & \text{if } \text{event} = \text{Job start} \\
    -1.0 & \text{if } \text{event} = \text{LO-job budget overrun} \\
    -2.0 & \text{if } \text{event} = \text{HI-job budget overrun} \\
    0.0 & \text{otherwise}
\end{cases}
\]

With rewards for events in place, the reward for an action is the sum of rewards for the events that occur between the application of the action and the next agent activation, i.e. the start of the next agent job. Thus the agent has to save the action and the state in an auxiliary structure to be queried in the following agent activation.

To support the learning process, the environment, more specifically the scheduler, needs to notify the agent of the occurrence of these events.

\subsection{DQN structure}

The agent uses a simple multi-layer feedforward neural network. The number of nodes of the input and of the ouput layers are as described above. The number of middle layers and the number of nodes per layer are hyperparameters that we have chosen as described below in Section~\ref{sec:evaluation}.

An activation function is applied to the output of all nodes of all but the last layer. Based on our experiments, detailed in Section~\ref{sec:evaluation}, a deployed agent always uses ReLU as its DQN network activation function.

%% file: environment.tex
\section{Environment Simulation}\label{sec:environment}

In order to train and evaluate the agent, we developed an environment simulator. Essentially, it comprises an event driven simulation engine, a set of event types and their handlers and an implementation of AMC. We briefly describe each of these components in the following paragraphs.

\paragraph{Event-driven simulation engine} This is a simple engine. It has an event queue, i.e. a priority queue with the pending events ordered in an order consistent to their time of occurrence. The engine just executes an endless loop, in which it removes the event at the head of the event queue, updates the current time to that of the event and invokes the respective handler.

\paragraph{Events and their handlers} There are 3 event types: job arrival, job completion and budget overrun. Each of these types is handled by its own handler, which takes the appropriate actions. For example, in the case of a job arrival, the handler samples the job execution time for the job, as described in Section~\ref{sec:task_execution_times}, adds the job to the queue of ready jobs and invokes the scheduler.  If the arriving job is scheduled to run, the handler will also schedule the respective job termination event assuming that the job will not be preempted. In LO-mode, the job termination event is either a job completion or a budget overrun, depending on the job execution time. In HI-mode, there are no budget overrun events: by assumption, the HI-WCET estimates are safe. In addition, if the job arrival causes the preemption of a job, the handler cancels the respective termination event.

Invocation of the agent is also triggered by the event handlers. Remember that the agent is the task with the lowest priority. Thus, when the agent job is selected by the scheduler to run, the event handler that called the scheduler will invoke the agent. However, it will not apply its output immediately. The output will be applied only upon the completion event of the agent job. Note that the agent task is a special task in that it is not assigned any budget: because it has the lowest priority, it cannot interfere with the execution of the other tasks. (For the sake of a uniform implementation, we assign the agent task the maximum value accepted by the type we use for the budget variables.)

Although, as we mentioned the order of the events in the event queue is consistent with the time of occurrence of the events, we use additional rules to break ties. In the case of events with the same time, we give priority to termination events. Note that the event queue has at most one termination event at any time: only the running process has a termination event, other jobs do not. Furthermore, in the case of job arrival events with the same time, we give priority to jobs with higher priority. The reason for these rules is simulation efficiency. Consider, for example, the beginning of a simulation. Before entering the event loop, the simulator initializes the event queue with a job arrival for each job at time 0. Assume that the first event in the queue was the arrival of the job with the lowest priority. Because the processor is idle at that time, that job would be scheduled to run, and its termination event added to the event queue. However, immediately after, when the simulator handles the next event in the queue, it will have to preempt that job, canceling the termination event of the running job, which did not actually run, because time has not advanced. By handling the arrival of the highest priority job first, the simulator will not have to undo any action when processing the other events that occur at the same time. It will just have to add the other jobs arriving at the same time to the ready queue.

\paragraph{Scheduler}

The scheduler is an implementation of AMC, which is a fixed priority-based scheduler with additional functionality to handle operation modes. Furthermore, to support our approach, it provides an interface to allow changing budgets at run time. At its core it has a queue of jobs that are ready to run, the \emph{ready queue}. The jobs in the ready queue are kept in decreasing order of their priority. This speeds up selecting the job to run, when the running job terminates.

The scheduler is invoked directly by the event handlers as described above, and takes the appropriate actions to handle each scheduling event. For example, in the case of a job arrival, if the processor is idle, the scheduler will remove the job at the head of the ready queue. If the processor is not idle, the scheduler compares the priorities of the running job and of the job at the head of the ready queue, and will select to run the one with higher priority. In the case of preemption, the preempted job is added to the ready queue. The scheduler has to return to the event handler information about its scheduling decisions, so that the event handler can take the appropriate actions, e.g. remove the termination event of the preempted job.

Our implementation of AMC is a variant of the algorithm described in \cite{baruah_response-time_2011} and summarized in Section~\ref{sec:AMC}. More specifically, it handles budget overruns differently. Whereas in \cite{baruah_response-time_2011}, there is a mode change upon every budget overrun, our implementation handles budget overruns by LO-jobs differently from those by HI-jobs. In the case of a budget overrun by a LO-job, the scheduler just cancels the "misbehaving" job. Indeed, this is enough to ensure that every other job, either LO- or HI-, will still meet its deadline, as long as it does not overrun its budget. On the other hand, in the case of a budget overrun by a HI-job, switching to HI-mode is necessary to ensure that that job can still meet its deadline. Thus the scheduler switches to HI-mode and removes all LO-jobs from the ready queue, so that they will not run while in HI-mode. We will refer to this AMC variant as AMC+.

Switching back to LO-mode occurs when the scheduler is notified of a job completion event and the ready queue is empty. As a result the event handler will schedule the arrival of LO-jobs that were either cancelled at the time of the LO- to HI-mode switch, or that have arrived while the system was in HI-mode.

%% file: evaluation.tex
\section{Experimental Evaluation}\label{sec:evaluation}

In this section, we present a simulation-based evaluation of the performance of the DQN-agent using a realistic automotive application~\cite{kramer_real_automotive}. Our evaluation aims to evaluate how the DQN-agent reduces budget overruns. Therefore, we use as metrics the number of HI-job budget overruns, which lead to mode changes, and LO-job budget overruns, which lead to LO-job kills.

\subsection{Task set generation}\label{sec:tasksets}

To generate the task sets we used the method presented in \cite{kramer_real_automotive}, which describes the characteristics of realistic automotive applications. These applications consist of components that contain runnables that are subject to scheduling. Runnables are mapped to tasks, which are scheduled by the operating system.

\subsubsection{Runnables} 

In \cite{kramer_real_automotive}, the authors provide a detailed timing characterization of the runnables of a typical automotive application.

We assume that all runnables are periodic. This differs from \cite{kramer_real_automotive} which states that about 15\% of the runnables are angle-synchronous, i.e. related to the crankshaft rotation in internal combustion engines. Thus, we have normalized the shares of the runnables per periods, as shown in Table~\ref{tbl:periods}. The second column is the share we used in our evaluation, whereas the third column is the share presented in \cite{kramer_real_automotive}.
\begin{table}
    \centering
    \begin{tabular}{|r|r|r|} \hline
     \textbf{Period} & \textbf{Share} & \textbf{Share (\cite{kramer_real_automotive})}\\ \hline
      1 ms   & \textbf{4\%} & 3\%\\
      2 ms   & \textbf{2\%} & 2\% \\
      5 ms   & \textbf{2\%} & 2\% \\
      10 ms   & \textbf{29\%} & 25\% \\
      20 ms   & \textbf{29\%} & 25\% \\
      50 ms   & \textbf{4\%} & 3\% \\
      100 ms   & \textbf{24\%} & 20\% \\
      200 ms   & \textbf{1\%} & 1\% \\
      1000 ms   & \textbf{5\%} & 4\% \\ \hline
     \end{tabular}
    \caption{Runnables period distribution}
    \label{tbl:periods}
\end{table}

The method described in \cite{kramer_real_automotive} is supposed to be applicable to applications of different complexity and to platforms of different capacity, thus the total number of runnables is not provided. We describe the approach we used to determine that number later in this section.

To determine the WCET of each runnable, first, we generate the average case execution time (ACET) of a runnable, according to the parameters of Table \ref{tbl:ACET}, reproduced from \cite{kramer_real_automotive}. This table provides the minimum, average and maximum ACET for each runnable period. To ensure that the average of the ACETs of all runnables with the same period is the one specified in this table, we have used UUniFast \cite{bini_uunifast_2005}. UUniFast allows to compute $n$ non-zero addends of a sum between 0 and 1 in such a way that the values of the addends are uniformly distributed between 0 and the value of the sum. Thus, we distributed the total processor utilization of the $r$ runnables with period $T$, $r \cdot ACET_{avg}/T$, by each runnable, according to UUniFast. To ensure that the minimum and maximum ACET bounds were respected, we discarded all distributions in which at least one runnable had an ACET outside the interval defined by those bounds.

\begin{table}
    \centering
    \begin{tabular}{|r|r|r|r|} \hline
        \textbf{Period} & \textbf{Min. ACET} & \textbf{Avg. ACET} &     \textbf{Max. ACET} \\ \hline
          1 ms & 0.34 $\mu$s & 5.00 $\mu$s & 30.11 $\mu$s \\
          2 ms & 0.32 $\mu$s & 4.20 $\mu$s & 40.69 $\mu$s \\
          5 ms & 0.36 $\mu$s & 11.04 $\mu$s & 83.36 $\mu$s \\
          10 ms & 0.21 $\mu$s & 10.09 $\mu$s & 309.87 $\mu$s \\
          20 ms & 0.25 $\mu$s & 8.74 $\mu$s & 291.42 $\mu$s \\
          50 ms & 0.29 $\mu$s & 17.56 $\mu$s & 92.98 $\mu$s \\
          100 ms & 0.21 $\mu$s & 10.53 $\mu$s & 420.43 $\mu$s \\
          200 ms & 0.22 $\mu$s & 2.56 $\mu$s & 21.95 $\mu$s \\
          1000 ms & 0.37 $\mu$s & 0.43 $\mu$s & 0.46 $\mu$s \\ \hline
    \end{tabular}
    \caption{Runnable Average Execution Times (from Table IV in \cite{kramer_real_automotive}}
    \label{tbl:ACET}
\end{table}

Once the ACET of a runnable is determined, its WCET is obtained by multiplying it by a factor that is uniformly sampled from an interval that depends on the period of the runnable. Table~\ref{tbl:BCET-WCET} shows the bounds of that factor for each period, in the 4th and 5th columns. The 2nd and 3rd columns have the bounds of the factor to determine the best case execution time (BCET).

\begin{table}
    \centering
    \begin{tabular}{|r|r|r|r|r|} \hline
     \textbf{Period} & $\mathbf{f_{min}}$ & $\mathbf{f_{max}}$ & $\mathbf{f_{min}}$ & $\mathbf{f_{max}}$\\ \hline
      1 ms   & 0.19 & 0.92 & 1.30 & 29.11\\ 
      2 ms   & 0.12 & 0.89 & 1.54 & 19.04 \\
      5 ms   & 0.17 & 0.94 & 1.13 & 18.44 \\
      10 ms   & 0.05 & 0.99 & 1.06 & 30.03 \\
      20 ms   & 0.11 & 0.98 & 1.06 & 15.61 \\
      50 ms   & 0.32 & 0.95 & 1.13 & 7.76 \\
      100 ms  & 0.09 & 0.99 & 1.02 & 8.88 \\
      200 ms   & 0.45 & 0.98 & 1.03 & 4.90 \\
      1000 ms   & 0.68 & 0.80 & 1.84 & 4.75 \\ \hline
     \end{tabular} 
    \caption{Factor for computation of BCET and WCET (from Table V in \cite{kramer_real_automotive})}
    \label{tbl:BCET-WCET}
\end{table}

\paragraph{Computing the parameters of the distribution of the execution time of each runnable} The  distribution of the execution time of each runnable can be approximated by a Weibull distribution \cite{kramer_real_automotive}. We use the ACET, BCET and WCET of each runnable to derive the parameters of the Weibull distribution of its execution time, which is used for computing the execution time of the tasks during simulation as described in Subsection\ref{sec:task_execution_times}. More specifically, we use the method described in \cite{mccombs_weibull_09} for the non-translated Weibull distribution. First, we compute the shape parameter, $k$, from the BCET and the WCET. We arbitrate that 10 ns\footnote{We use 10 ns, because that is the time resolution of our simulator.} and WCET-BCET are the outputs of the quantile function, i.e. the inverse of the cumulative probability function, for probabilities 0.00001 and 0.99999, respectively. Second, we use the shape factor and the ACET-BCET, i.e. the average rather than the mode as in \cite{mccombs_weibull_09}, to compute the scale parameter. Finally, we set the location parameter to the BCET of the runnable. 

\paragraph{Parameters not provided by Kramer et al.}
To determine the criticality level of each runnable, given that \cite{kramer_real_automotive} does not provide any guidance,  we have designed our experiments so that the actual value is not that relevant as we discuss below. So, we use a 50/50 distribution between the two modes.

\begin{figure}
    \centering
    \includegraphics[width=\linewidth]{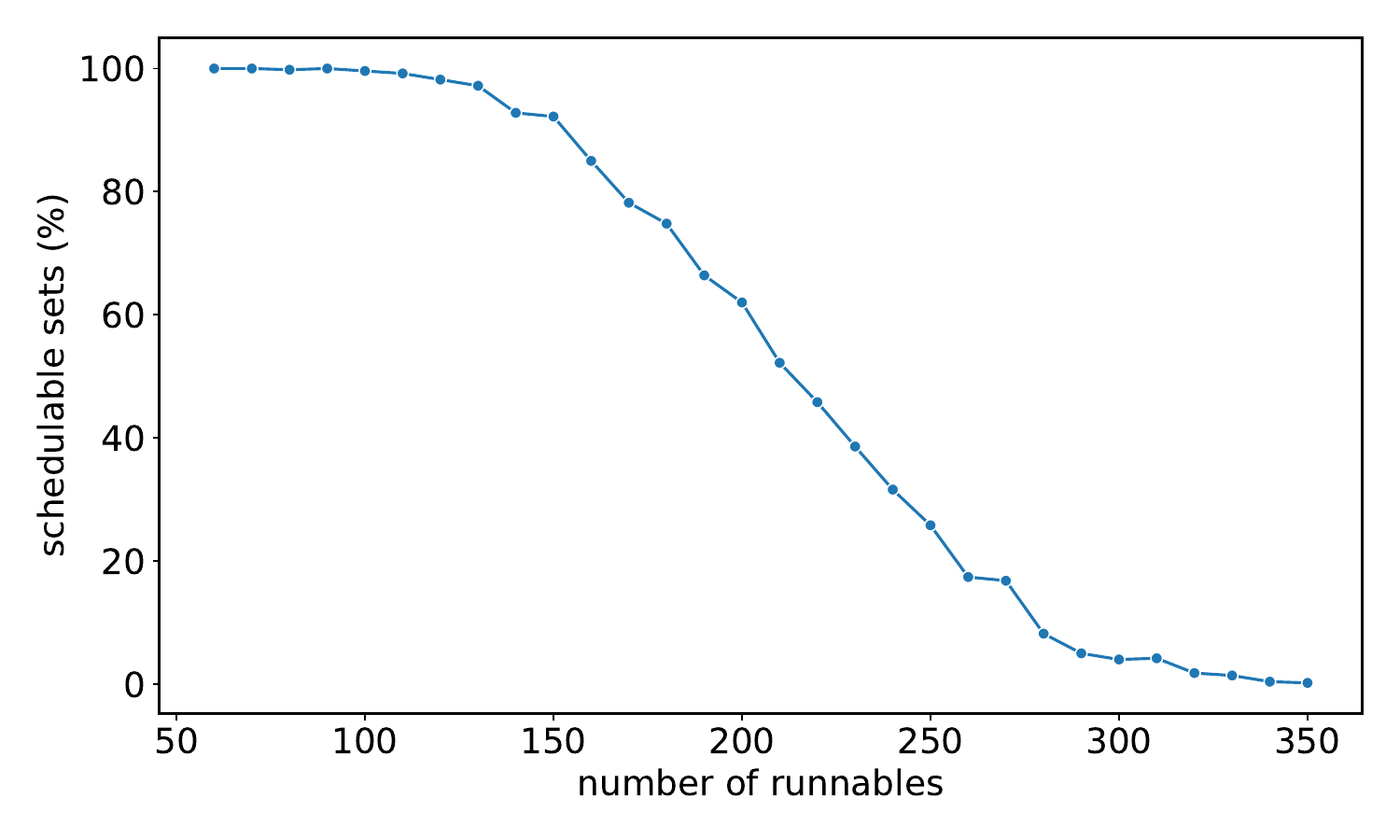}
    \caption{Percentage of guaranteed schedulable sets depending on the number of total runnables.}
    \label{fig:interval-of-interest}
\end{figure}

The number of runnables in the system is determined empirically, to better evaluate the efficacy of the proposed approach. Typically, the percentage of schedulable mixed criticality task sets using AMC-rtb follows a well defined pattern: for low loads almost all task sets are schedulable, whereas for high loads almost no task set is schedulable. In between these two regions, there is a region where the schedulability drops with a fairly constant slope. In the transition between regions, we observe a more or less smooth variation in the slope. We believe that adjusting the budgets has a higher potential in the region with a non-zero slope, regardless of the actual parameters of the task set, including the ratio betwen HI and LO runnables. In our set up, we can vary the load by varying the number of runnables. Therefore, to determine the number of runnables of the middle region, i.e. the one with a non-zero slope, we have generated 10 task sets with a given number of runnables, using the method described in Sections \ref{sec:tasks} and \ref{sec:task_execution_times}, and executed the AMC-rtb test for each of these sets. We started with a 20 runnables and initially incremented this number with a step of 30. Once we located the interval where the schedulable ratio decreases, we reduced the step to 10. The results obtained are shown in Fig.~\ref{fig:interval-of-interest}. Based on these results we used the following set of values for the number of runnables: 150 and 250.

\subsubsection{Application Tasks}\label{sec:tasks}

In automotive applications, usually there is only one task per period, although for reasons of separation and distribution there can be more than one \cite{kramer_real_automotive}. 

Because we are using a mixed criticality model, we create one task per period per mode, and map all runnables with this period and mode to the respective task. Thus, in total there are at most 18 tasks. Since the characteristics of the runnables are random as described above, for periods with a small share of runnables, some tasks may be assigned 0 runnables, and therefore the number of tasks may be lower.

We set the deadline of each task to its period, according to \cite{kramer_real_automotive}.

To compute the WCET in HI-mode of a HI-task, we add the WCET of each runnable mapped to the task.

To choose the WCET in LO-mode for both HI- and LO-tasks, we use the quantile function of the execution time of each task. The  probabilities chosen as input of the quantile function depend  on the task period and criticality, as shown in Table~\ref{tbl:quantiles}. Given a task, we determine the value of the quantile function for the input probability experimentally. More specifically, we sample the execution time of 1000 jobs of the task as described in the next section. These values are then sorted in increasing order of their values and we choose the value of the element whose position in the sorted list is $p*1000$, where $p$ is the probability drawn from Table~\ref{tbl:quantiles}. 

The rationale for the values in this table is to reduce both the probability and the frequency of events  that may cause QoS degradation. Indeed, the outputs of the quantile functions are budgets in LO-mode, and when a job exceeds its budget either the job is cancelled or there is a mode switch canceling all the jobs of LO-tasks. Thus, we use higher input probabilities for HI-tasks than for LO-tasks. Finally, to reduce the frequency of these events, the shorter the period of the task the higher the input probability. Although this approach is sound,  the probability values in Table~\ref{tbl:quantiles} are much lower than those that would be used since budget overruns should be very rare events~\cite{burns13_more_pratical_model_for_MCS_WMC}. The reason for this choice was to make it possible to evaluate our approach in a reasonable time. If we used more realistic values, rather than simulate the system for 1000 seconds, we would have to simulate it for a much longer interval.

\begin{table}
    \centering
\begin{tabular}{|r|r|r|} \hline
\textbf{Period (/ms)} & \textbf{LO-task} & \textbf{HI-task} \\ \hline
1 & 0.75 & 0.8 \\
2 & 0.75 & 0.8 \\
5 & 0.75 & 0.80 \\
10 & 0.67 & 0.75 \\
20 & 0.67 & 0.75 \\
50 & 0.67 & 0.75 \\
100 & 0.5 & 0.67 \\
200 & 0.5 & 0.67 \\
1000 & 0.5 & 0.67 \\ \hline
\end{tabular}
    \caption{Input probabilities for computation of L-WCET}
    \label{tbl:quantiles}
\end{table}

\subsubsection{Agent Task}

As already mentioned, the agent task runs as a HI-task with the lowest priority among all tasks, and its execution must also be simulated. We assign no budget to the agent task, because it runs only when no application task is ready to run; this ensures that no agent job is killed. Furthermore, it is implemented as a sporadic task with a minimum inter-arrival time of 10 ms; at any time, there is at most one agent task ready to execute. The execution time of each agent job is sampled from a Weibull distribution at arrival time.

\begin{figure}
    \centering
    \includegraphics[width=0.7\linewidth]{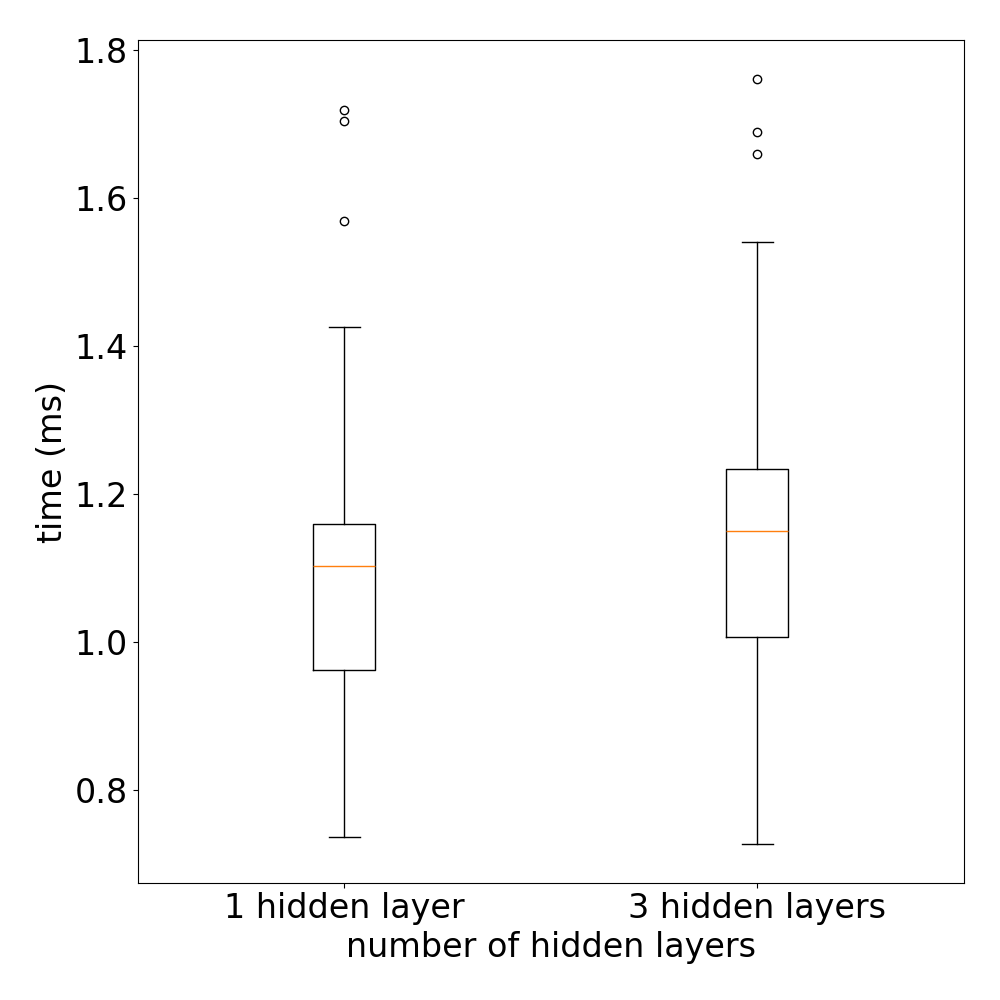}
    \caption{Agent execution times on a Raspberry Pi 4 Model B.}
    \label{fig:activation_times_rasp}
\end{figure}

 We derived the parameters of the Weibull distribution as described above, using 750 $\mu\text{s}$, 1200 $\mu\text{s}$, 760 $\mu\text{s}$ and 2000 $\mu\text{s}$, for the BCET, ACET and 0.000001 and 0.99999 quantiles, respectively. These values were determined by running the agent 1000 times and measuring its execution time on each of these runs on a Raspberry Pi Model B, an embedded systems platform. The distribution of the measured times is shown in Fig.~\ref{fig:activation_times_rasp}, for two different DQN models (see Subsection~\ref{sec:hyperparameters}).

\subsection{Generation of execution time of simulated tasks}\label{sec:task_execution_times} At run-time, the simulator determines the execution time of an arriving job based on the properties of the respective task.

Given that tasks are composed of runnables, we sample the execution time of each runnable from the respective Weibull distribution and add all the sampled execution times to get the job execution time.

\subsection{Deep Q-Network Hyperparameters}\label{sec:hyperparameters}

Since we know the task set before deployment, the generated DQN agent can be engineered to perform best for that specific task set. In other words, there is no need for our model to generalize, and thus, there is room for using different sets of hyperparameters for each agent.

Based on common practices, we started with the set of hyperparameters presented in Table ~\ref{tab:hyperparameter_tuning_values}. The number of hidden layers and their sizes of both the policy and the target  networks, the replay memory sample batch size and the activation functions vary, whereas the others are fixed.
Table~\ref{tab:hyperparameter_tuning_values} represents the size of the hidden layers as arrays. E.g.~the array $[n, n/2]$ denotes a DQN with two hidden layers, the first of which has $n$ nodes and the second $n/2$ nodes, where $n$ is the task set size.

\begin{table}[!h]
\centering
\begin{tabular}{|l|l|}
\hline
\textbf{Hyperparameter} & \textbf{Values Space} \\ \hline
maximum memory size & 200 \\ \hline
minimum memory size & 20 \\ \hline
gamma & 0.99 \\ \hline
network update frequency & 5 \\ \hline
learning rate & \(5 \times 10^{-5}\) \\ \hline
hidden layers size & \([n/2], [n, n/2], [n, n/2, n/4]\) \\ \hline
sample batch size & 3, 6, 12 \\ \hline
network activation function & sigmoid, relu, tanh \\ \hline
\end{tabular}
\caption{The search space of hyperparameter values used for hyperparameter tuning. $n$ is the number of tasks in the task set.}
\label{tab:hyperparameter_tuning_values}
\end{table}

\begin{figure}[h!]
    \centering
    \begin{subfigure}{0.29\linewidth}
        \includegraphics[width=\linewidth]{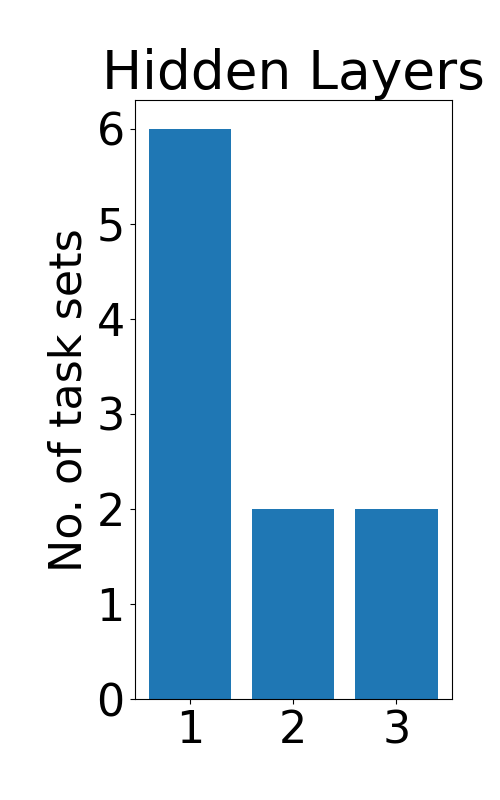}
    \end{subfigure}\hfill
    \begin{subfigure}{0.29\linewidth}
        \includegraphics[width=\linewidth]{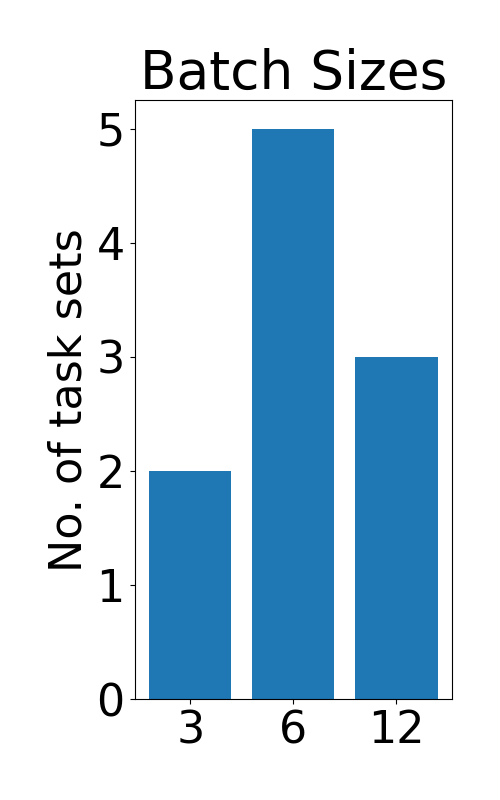}
    \end{subfigure}\hfill
    \begin{subfigure}{0.41\linewidth}
        \includegraphics[width=\linewidth]{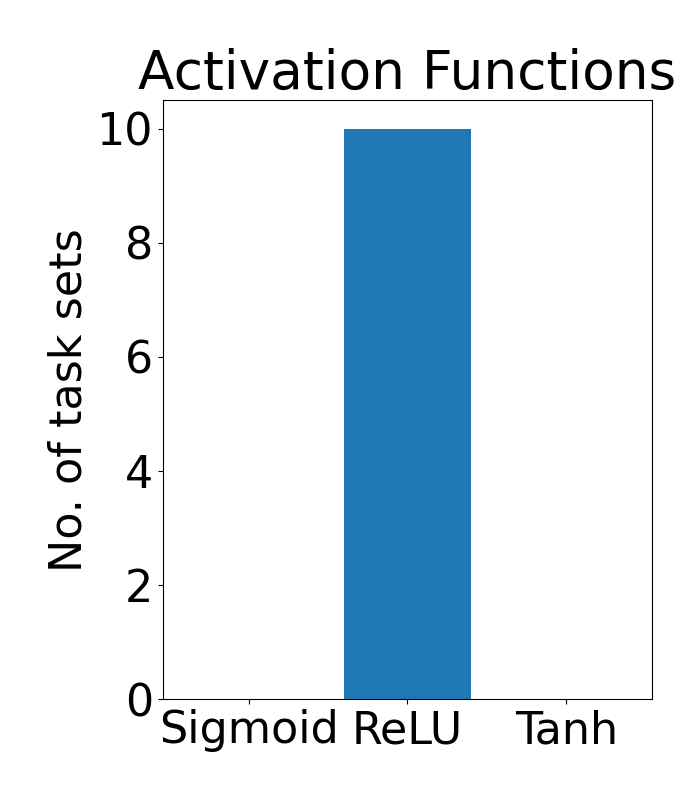}
    \end{subfigure}
    \caption{Number of task sets for which the hyperparameter values that we varied performed best, across simulations of 10 task sets with 150 runnables each.}
    \label{fig:hyperparameter_best}
\end{figure}

To reduce the time required for our experimental evaluation, we performed an initial analysis of the impact of the chosen factors on the performance of the DQN agent. More specifically, we conducted a full-factorial experiment in which we trained and tested different models, one for each combination of the values of the different factors, using 10 different task sets with 150 runnables each. Figure ~\ref{fig:hyperparameter_best} illustrates how frequently the different hyperparameter values lead to the best-performing model for each task set. ReLU is the activation function that always performs best, i.e., yields the most reward in test simulations. For the other two hyperparameters, no value dominates the others. Therefore, we used ReLU as an activation function in our evaluation experiments and varied the other two hyperparameters.

\subsection{Evaluation Results}\label{sec:results}

To evaluate our approach, we randomly generated 100 task sets for each number of runnables, as described in Section~\ref{sec:tasksets} and \ref{sec:task_execution_times}. 

All 9 DQN models were trained for 100 seconds of simulated time. 

For each task set, we simulated its execution during 1000 simulated seconds, using AMC+ and AMC+ with each of the 9 DQN models. The values reported for our approach are the best among all 9 models trained. 

\begin{figure}
    \centering
    \includegraphics[width=0.8\linewidth]{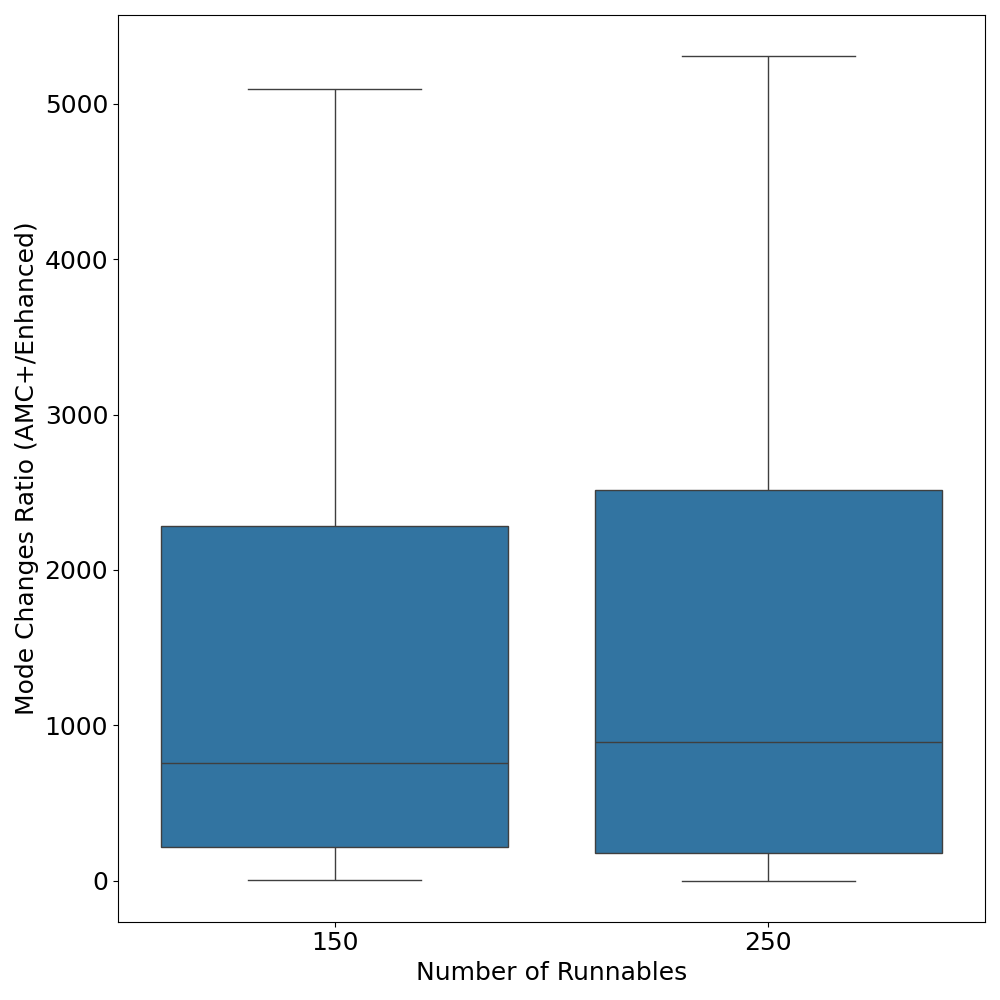}
    \caption{Improvement observed in mode changes with and without an agent across 100 task sets with 150 and 250 runnables, excluding outliers.}
    \label{fig:task_mode_changes_cmp_relative}
\end{figure}

\begin{figure}
    \centering
    \includegraphics[width=0.8\linewidth]{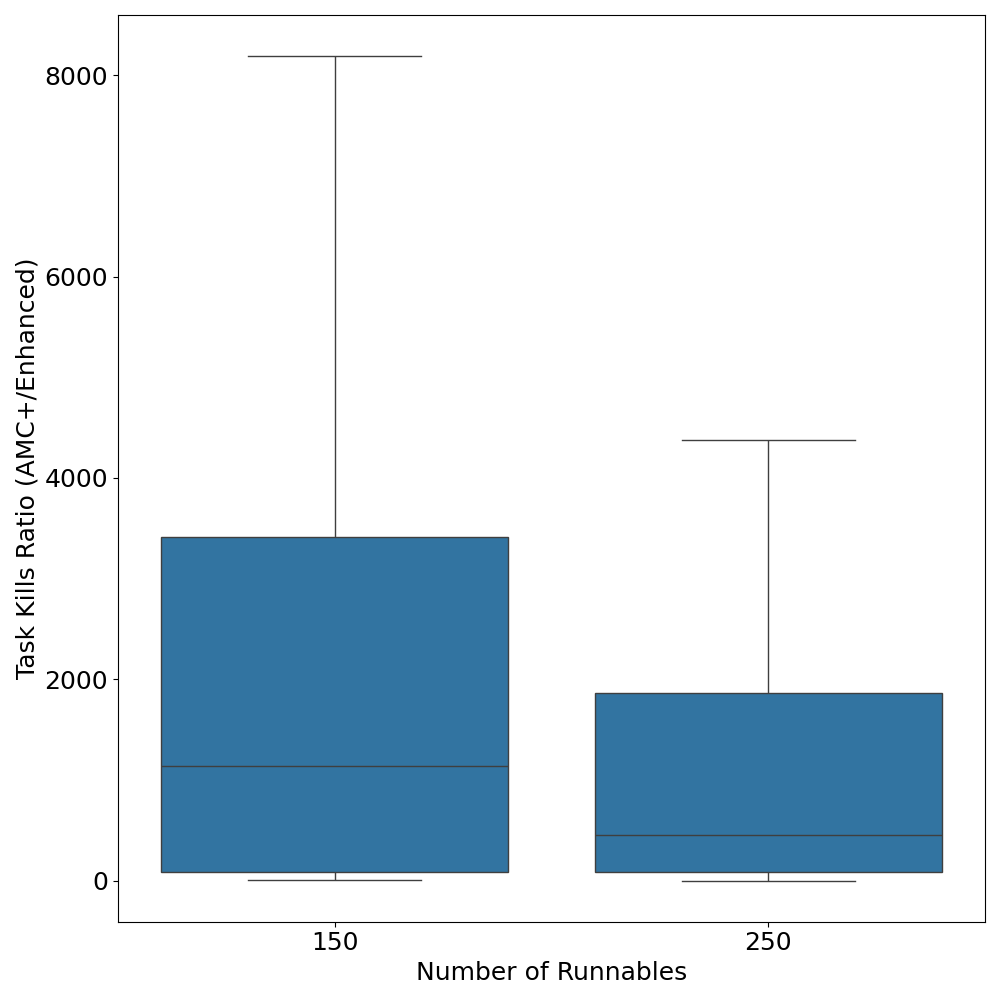}
    \caption{Improvement observed in job cancellations with and without an agent across 100 task sets with 150 and 250 runnables, excluding outliers.}
    \label{fig:task_kills_cmp_relative}
\end{figure}

Figures~\ref{fig:task_mode_changes_cmp_relative} and \ref{fig:task_kills_cmp_relative} show the distribution of the ratio between each metric using AMC+ and the AMC+ enhanced with the DRL agent. For easier readability this information is provided also in Tables~\ref{tbl:mode_changes_cmp_relative} and \ref{tbl:task_kills_cmp_relative}.

We use the ratio between each metric using AMC+ and the AMC+ enhanced with the DRL agent, because the absolute values of these metrics are dependent on the probabilities used to set the budgets in LO-mode, see Table~\ref{tbl:quantiles}.

On average, for each configuration, for both metrics, the ratio between the value without the DQN agent and the value with the DQN agent is larger than 1.5, as we would expect, because 1.5 was the worst case among the 10 task sets we showed. Clearly, the DQN agent can reduce significantly the number of budget overruns.

Comparing the results for both configurations, i.e. 150 and 250 runnables, the results are mixed. In principle, we would expect the agent to have more difficulties to improve in configurations with 250 runnables, because the load is higher than for task sets with 150 runnables. This intuition is confirmed by the minimum values shown in the Tables~\ref{tbl:mode_changes_cmp_relative} and \ref{tbl:task_kills_cmp_relative}: they are smaller for task sets with 250 runnables than for task sets with 150 runnables.

\begin{table}[h!]
    \centering
    \caption{Quantile comparison of mode changes improvement ratio with and without an agent across 100 task sets with 150 and 250 runnables.}
    \begin{tabular}{lrrrrr}
    \toprule
    \textbf{Runnables} & \textbf{min} & \textbf{25\%} & \textbf{50\%} & \textbf{75\%} & \textbf{max} \\
    \midrule
    150 & 4.4 & 215.6 & 757.3 & 2285.8 & 23907.6 \\
    250 &      2.3 &     181.6 &    892.8 &   2513.1 &  41928.0    \\
    \bottomrule
    \end{tabular}
    \label{tbl:mode_changes_cmp_relative}
\end{table}

\begin{table}[h!]
    \centering
    \caption{Quantile comparison of job cancellations improvement ratio with and without an agent across 100 task sets with 150 and 250 runnables.}
    \begin{tabular}{lrrrrr}
    \toprule
    \textbf{Runnables} & \textbf{min} & \textbf{25\%} & \textbf{50\%} & \textbf{75\%} & \textbf{max} \\
    \midrule
    150 & 2.2 & 81.1 & 1140.6 & 3415.7 & 100456.0 \\
    250 &      1.5 &     89.8 &    450.1 &   1866.1 &  67011.0    \\
    \bottomrule
    \end{tabular}
    \label{tbl:task_kills_cmp_relative}
\end{table}

%% file: conclusions.tex
\section{Conclusions and Future Work}\label{sec:conclusion}

We developed a DQN agent that adjusts at run-time the budgets of all tasks, to reduce the number of budget overruns in LO-mode so as to reduce the number of LO-job cancellations in AMC scheduling, which can affect the quality of service of deployed systems. The actions of the agent are consistent with the AMC-rtb analysis assumed to be carried out at design time.  

To evaluate our approach, we simulated a realistic automotive application~\cite{kramer_real_automotive} and compared the number of LO-job cancellations and the number of mode changes of an AMC+ without and with the agent.

Our results show that the DQN agent can significantly reduce both the number of LO-job cancellations and mode changes. 

We believe that our approach can be applied in similar real-time resource management problems.

There is however an open question. In our experiments, we have set the budgets of tasks in LO-mode to a relatively low value, so as to generate sufficiently frequent budget overruns to evaluate our agent in acceptable time. But, this also reduces the time to train the agent. So the open question is whether it is feasible to train and or test the agent when the budgets are set using common practice.